\documentclass[final,3p,times,twocolumn]{elsarticle}

\usepackage{amssymb}
\usepackage{amsmath}
\usepackage{xcolor}
\usepackage{textcomp}

\biboptions{numbers,sort&compress}

\usepackage{hyperref}

\journal{Journal of Magnetism and Magnetic Materials}

\begin{document}

\begin{frontmatter}

\cortext[cor1]{jgpark10@snu.ac.kr}

\title{Van der Waals Antiferromagnets: From Early Discoveries to Future Directions in the 2D Limit} 
\author{Rahul Kumar, Je-Geun Park\corref{cor1}} 

\affiliation{organization={Department of Physics and Astronomy, Seoul National University},
            city={Seoul},
            postcode={08826}, 
            country={South Korea}}

\begin{abstract}

The emergence of a long-range magnetic order in the atomically thin, two-dimensional (2D) limit has long remained a fundamental question in condensed matter physics. The advent of exfoliable van der Waals (vdW) materials, particularly transition-metal phosphorus trisulfides ($TM$PS$_3$; $TM$ = Fe, Ni, and Mn), provided the first experimental access to this regime and established a foundational platform for investigating 2D magnetism. The 2016 experimental demonstrations of intrinsic magnetism in monolayer FePS$_3$ provided a platform to test key aspects of 2D Ising criticality in the true 2D limit. It was followed by a rapid growth resulting in a wealth of emergent phenomena arising from the interplay of low-dimensional magnetism and quantum materials. We begin this review with the historical development of vdW antiferromagnets and highlight the key physical insights gained over the past decade. We finish with emerging opportunities in which vdW antiferromagnets can serve as versatile platforms for exploring low-dimensional magnetism and its interplay with other quantum degrees of freedom.

\end{abstract}


\begin{highlights}
\item Overview of vdW antiferromagnets evolution from 2016 and the future prospects.
\item Understanding the crystal structure and magnetism of $TM$PS$_3$ family members, key players of the vdW antiferromagnetic family.
\item Challenges faced while characterizing the vdW magnets in 2D limit and recent advances.
\item Review of the new physics unlocked by vdW magnets and their potential use in future spintronics.
\end{highlights}


\end{frontmatter}



\section{Introduction — Origin of the Question}
\label{sec1}
Throughout the history of science, materials have often played a central role in advancing the frontiers. New materials open the sealed box of curiosity and imagination, allowing people to ask questions that were previously unimaginable. The discovery of graphene \cite{novoselov2004electric} is arguably the watershed moment in 21$^{st}$ century physics that brought the now-vibrant field of 2D materials into being. Subsequently, the question and the idea behind vdW magnets were a natural consequence of these developments in the research of 2D materials \cite{li2014half, sivadas2015magnetic, park2016opportunities}. 

As a separate article is dedicated to vdW ferromagnets, we focus exclusively on vdW antiferromagnets here. We encourage readers to consult both articles in parallel to gain a comprehensive perspective on recent developments in the broader landscape of vdW magnets. This article focuses on the historical development and experimental foundations of vdW antiferromagnets.

The isolation of graphene via a simple mechanical exfoliation method assisted by Scotch tape in 2004 \cite{novoselov2004electric} demonstrated that atomically thin (truly 2D) materials could be extracted from vdW crystals while retaining their remarkable properties. This breakthrough transformed condensed matter physics and sparked intense efforts to isolate other layered materials such as transition-metal dichalcogenides and hexagonal boron nitride  \cite{novoselov2005two,radisavljevic2011single}. Because of the simplicity of this approach, scotch tape exfoliation is the widely used technique for peeling a vdW material to a monolayer thickness, which has led to an exponentially growing catalog of vdW materials. Yet, despite this explosive growth, the magnetic analog of graphene received virtually no attention until 2010, when a pivotal question emerged: Does a magnetic vdW crystal analogous to graphene exist?

\begin{figure*}
\centering
\includegraphics[width=0.8\linewidth]{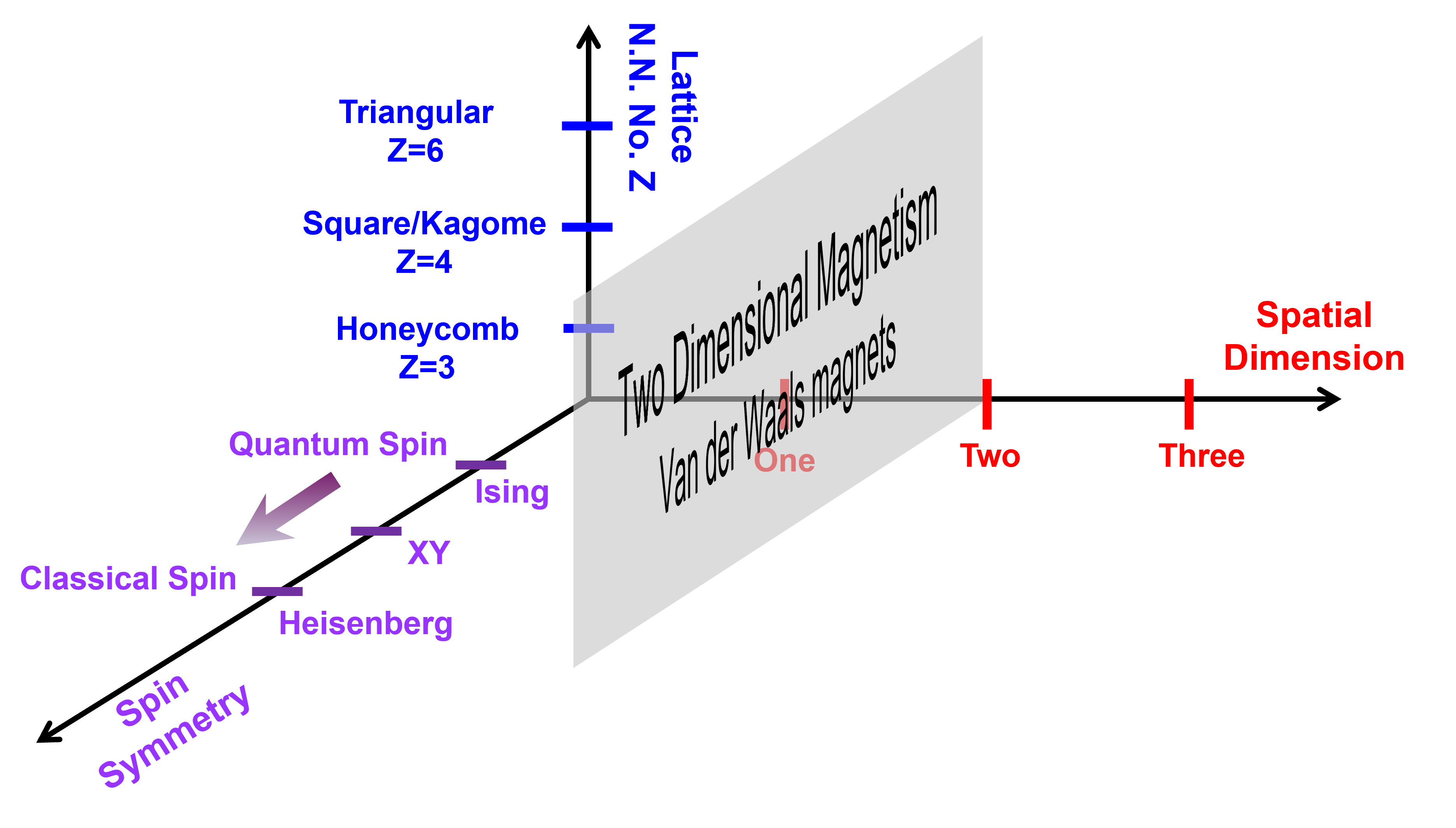}
\caption{Conceptual landscape of 2D magnets within the three-dimensional space of spatial dimensionality, lattice geometry and spin symmetry.}\label{fig1}
\end{figure*}

This material would not only be conceptually compelling, but it would also allow experimental tests of the canonical 2D spin models—Ising \cite{onsager1944crystal}, XY \cite{berezinskii1971destruction,kosterlitz1973ordering}, and Heisenberg \cite{mermin1966absence}—in atomically thin magnetic crystals. These materials could serve as valuable testbeds for investigating 2D magnetism in a range of spin-lattice geometries, including honeycomb \cite{lee2016ising}, triangular \cite{park2022field}, and Kagome \cite{ortiz2019new} structures. Figure \ref{fig1} conceptualizes the 2D magnets within this landscape. Our initial experimental attempts were made using honeycomb oxides such as Li$_2$MnO$_3$ \cite{lee2012antiferromagnetic}, which failed due to strong covalent bonding that prevented easy exfoliation.

\begin{figure*}
\centering
\includegraphics[width=0.8\linewidth]{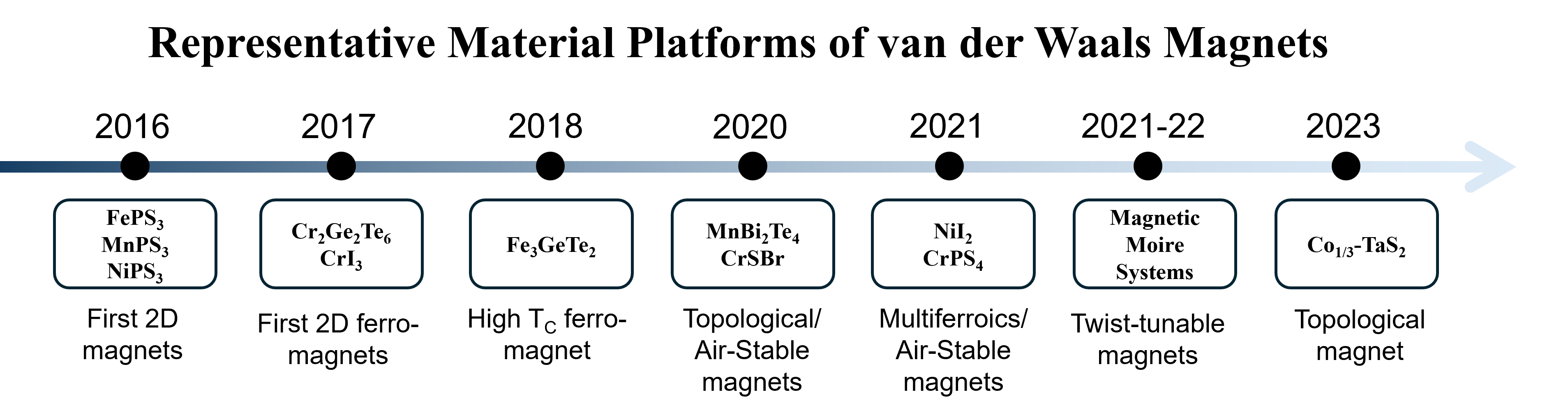}
\caption{Chronological development of van der Waals magnetic material platforms 
(2016--2023). The timeline traces major experimental discoveries and their 
characterization, beginning with the first two-dimensional magnets (FePS$_3$, MnPS$_3$, NiPS$_3$) in 2016, followed by the first ferromagnetic systems (Cr$_2$Ge$_2$Te$_6$, CrI$_3$) in 2017, and progressing through successive breakthroughs including high-temperature ferromagnets (Fe$_3$GeTe$_2$, 2018), topological and air-stable antiferromagnets (MnBi$_2$Te$_4$, CrSBr, 2020--2021), multiferroic and air-stable systems (NiI$_2$, CrPS$_4$, 2021), twist-tunable magnets (2021--2022), and topological magnets (Co$_{1/3}$TaS$_2$, 2023).}\label{fig2}
\end{figure*}

The breakthrough came through a systematic literature review, particularly revisiting Brec's earlier characterization of $TM$PS$_3$ \cite{brec1986review}. Unlike oxides, $TM$PS$_3$ has a weak interlayer bonding, essential for exfoliation, combined with a perfect honeycomb magnetic lattice \cite{du2016weak}. Another interesting point is that the $TM$PS$_3$ family demonstrated remarkable chemical flexibility with various 3$d$ transition-metal ions (Fe$^{2+}$, Ni$^{2+}$, Mn$^{2+}$, etc.) while maintaining a similar chemical structure. Most importantly, it realizes the three fundamental magnetic models: Ising, XY, and Heisenberg. Encouraged by the versatility of this family, Park {\it et al.} synthesized bulk crystals and demonstrated easy exfoliation down to monolayer FePS$_3$ \cite{lee2016ising}, NiPS$_3$ \cite{kuo2016exfoliation}, and MnPS$_3$ \cite{lee2016tunneling}. As the original photos show, it was immediately clear that the crystals could be peeled into thin flakes because of their low cleavage energies. The (re)discovery of this family of materials prompted a decade-long research program on vdW magnets that continues to expand today. This decade-long effort has made 2D magnetism a vibrant experimental discipline. Figure \ref{fig2} provides an overview of the representative vdW magnetic platforms discovered since 2016. This review complements recent comprehensive treatments by focusing on materials trends and experimental phenomenology rather than exhaustive theoretical formalism.

\section{Structural and Magnetic Properties of \texorpdfstring{$TM$PS$_3$}{TMPS3}}
\label{sec2}

The $TM$PS$_3$ family is a representative material platform, whose physical properties define the essential features and conceptual framework of 2D magnetism. Therefore, a comprehensive understanding of the structural and magnetic characteristics of the $TM$PS$_3$ family is necessary before delving into the details of vdW magnetism.

\subsection{Crystal structure and the origin of single-ion anisotropy}

The $TM$PS$_3$ family is distinguished by weak vdW interlayer interactions, enabling reliable mechanical exfoliation down to the monolayer limit. The magnetic ions form a honeycomb lattice within the crystallographic $ab$ plane, with a relatively large separation along the $c$ axis \cite{brec1986review,ouvrard1985structural}. This quasi-two-dimensional crystal structure, combined with weak interlayer bonding, makes $TM$PS$_3$ an ideal platform for investigating magnetic phenomena in the strict 2D limit.

Each transition-metal ion occupies the center of a sulfur octahedron ($TM$S$_6$ coordination polyhedron). The crystallographic distortion reduces the local point-group symmetry from cubic ($O_h$) to trigonal ($D_{3d}$), leading to a preferred magnetic easy axis or plane. Thus, the magnetic properties of $TM$PS$_3$ are governed by a delicate interplay between the crystal electric field and the spin-orbit coupling effects \cite{joy1992magnetism}. The resulting single-ion anisotropies can be further affected by the electronic configuration of the transition-metal cation, particularly when the $d$-electron configuration retains unquenched orbital angular momentum \cite{joy1992magnetism,kim2021magnetic}.

\subsection{Natural Realization of Canonical 2D Spin Models}
A remarkable feature of the $TM$PS$_3$ family is that different transition-metal cations, despite maintaining an identical crystal structure, exhibit fundamentally different magnetic anisotropies. This flexibility of magnetic symmetry with the same structure is not common among magnetic materials and is an important feature of $TM$PS$_3$. Crucially, this feature enables direct experimental access to the canonical 2D spin models: Ising, XY, and Heisenberg. This advantage of the $TM$PS$_3$ platform provides an huge opportunity to explore both dimensionality and symmetry in 2D magnetism, which can control magnetism independently.

\subsubsection{\texorpdfstring{FePS$_3$ (Ising like)}{FePS3 (Ising like)}}
The Fe$^{2+}$ cation ($3d^6$ configuration) has an unquenched orbital angular momentum ($^5T_{2g}$ state, $L \approx 1$, $S$ = 2). With a spin-orbit interaction, this leads to an unusually ``giant'' easy-axis anisotropy ($\sim$ 20~meV) \cite{lee2023giant,lanccon2016magnetic}, with the magnetic moments pointing strictly perpendicular to the honeycomb layers. The resulting magnetic configuration is effectively equivalent to the canonical 2D Ising model. 

\subsubsection{\texorpdfstring{NiPS$_3$ (XY like)}{NiPS3 (XY like)}}
In contrast, the Ni$^{2+}$ cation ($3d^8$ configuration) has a quenched orbital moment ($^3A_{2g}$ $L = 0$, $S$ = 1), such that single-ion anisotropy arises only from weaker second-order perturbative effects mediated by spin-orbit coupling of excited crystal-field states \cite{lee2024imaging,afanasiev2021controlling}. This weak mechanism generates a moderate easy-plane anisotropy that confines the magnetic moments to the honeycomb $ab$ plane with small canting. The resulting magnetic system is well-described by the 2D XXZ model with easy-plane anisotropy, commonly referred to as the 2D XY model in the limit of weak out-of-plane anisotropy \cite{coak2019tuning,wang2018new,scheie2023spin}.

\subsubsection{\texorpdfstring{MnPS$_3$ (Heisenberg like)}{MnPS3 (Heisenberg like)}}
\begin{figure*}[t]
\centering
\includegraphics[width=0.8\linewidth]{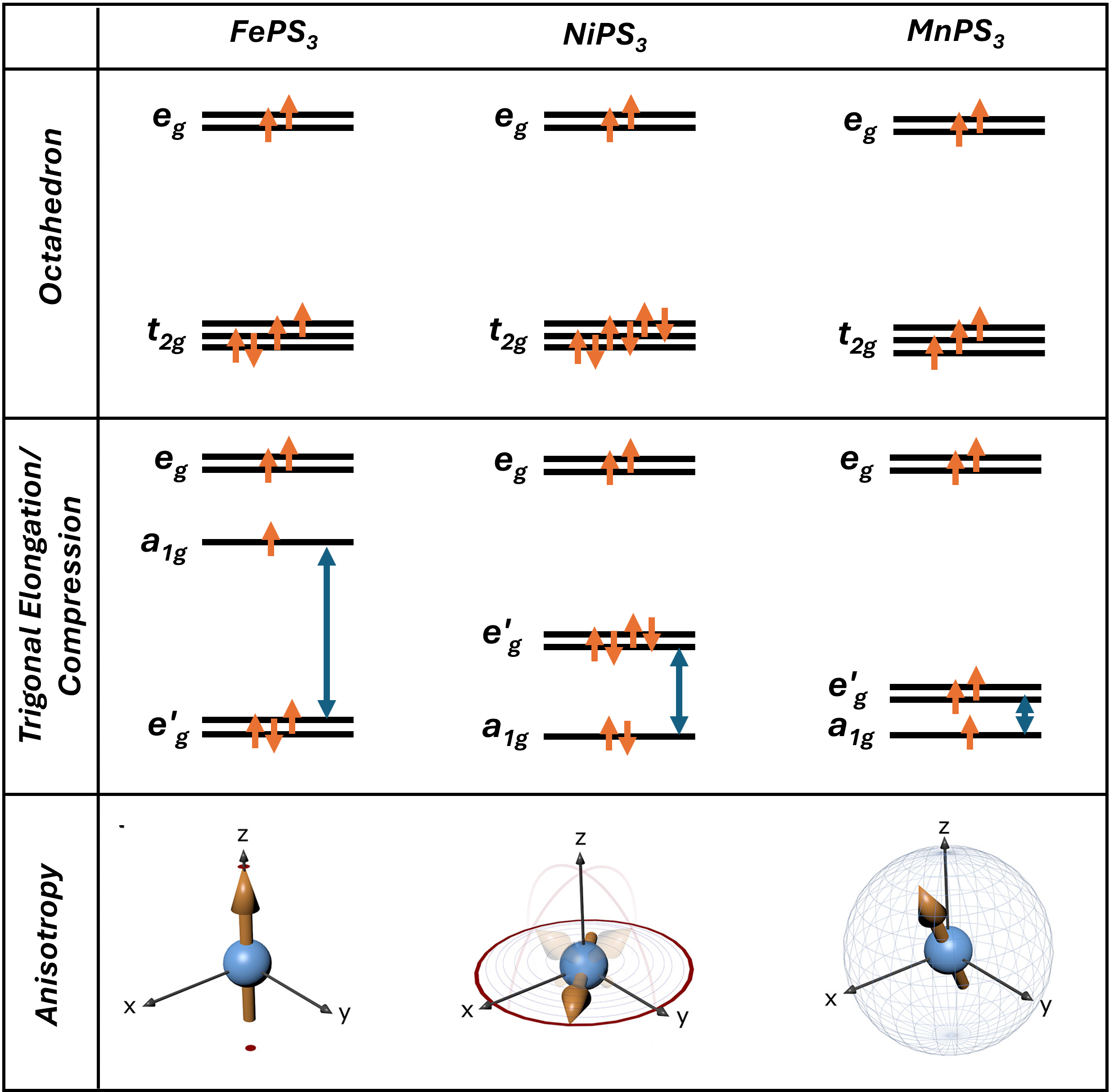}
\caption{Crystal field splitting of 3$d$ orbitals and resulting spin anisotropy in transition metal phosphorous trichalcogenides. \textit{Top row:} Octahedral crystal field splitting ($e_g$ and $t_{2g}$ orbitals) for FePS$_3$, NiPS$_3$, and MnPS$_3$. \textit{Middle row:} Trigonal elongation or compression of the octahedral environment, leading to further splitting of the $d$-orbitals into $e'_g$ and $a_{1g}$ states. The blue vertical arrows qualitatively illustrate the splitting. \textit{Bottom row:} Resulting magnetic anisotropy and spin models. The bottom row illustrations are adapted from \cite{wang2022magnetic}, Copyright \textcopyright 2022 The Authors.}\label{fig3}
\end{figure*}

The Mn$^{2+}$ cation ($3d^5$ configuration) features a half-filled $d$ shell with zero orbital angular momentum ($^6A_{1g}$ $L = 0$, $S$ = 5/2), rendering both first- and second-order spin-orbit coupling effects negligible. The resulting magnetic anisotropy is essentially zero, allowing the magnetic moments to point in any spatial direction with equal energy. This near-zero anisotropy \cite{wildes1998spin,olsen2021magnetic} allows MnPS$_3$ to approximate the isotropic 2D Heisenberg model, wherein spin interactions depend only on the relative orientation of neighboring spins rather than their absolute direction. MnPS$_3$ completes the trio of canonical 2D spin models realized within the $TM$PS$_3$ family. A summary of the 3$d$ orbital splitting and the resulting magnetic anisotropy for all three compounds is shown in Fig.~\ref{fig3}.

\vspace{5mm}
These properties make $TM$PS$_3$ one of the few material families where 2D spin models can be experimentally accessed in structurally similar compounds. However, probing these properties in genuinely 2D regimes (monolayer or few-layer) poses a serious experimental challenge. Due to the antiferromagnetic ground state, the total magnetic signal vanishes, except for small canted ferromagnetic moments. This limitation required the development of alternative experimental approaches specifically tailored to probe the antiferromagnetism of $TM$PS$_3$ in the strict 2D limit. This challenge has driven substantial early innovation in the field and will be discussed in subsequent sections.

\section{Overcoming experimental challenges (2012-14)}
Exfoliation of the $TM$PS$_3$ samples introduced additional experimental challenges, including sample handling and the study of magnetism in the truly 2D limit. The sample-handling challenges include charge compensation on the surface and stability under ambient conditions. Given the micrometer-scale lateral dimensions and atomic-scale thicknesses of exfoliated flakes, the available sample volume is too small for conventional bulk characterization techniques such as neutron scattering, superconducting quantum interference device (SQUID) magnetometry, and muon spin relaxation ($\mu$SR). Hence, to study 2D magnetism, a methodological transformation was required, specifically in the field of optical and electrical probes. The following subsections outline the key experimental challenges and the innovative strategies developed to overcome them.

\subsection{Initial sample-dependent challenges}
Unlike oxide magnets, vdW magnets, consisting of chalcogenides and halides, are always more sensitive to environmental conditions, which makes proper handling crucial. Bulk flakes of the $TM$PS$_3$ family are reasonably stable in air, but prolonged exposure (a few hours) to nonideal conditions such as air can easily degrade or damage exfoliated flakes, especially thin ones. Such effects were already reported in the first successful exfoliation study of NiPS$_3$ \cite{kuo2016exfoliation}. Exposure of NiPS$_3$ thin flakes to ambient air led to a pronounced reduction in optical contrast and Raman intensities over the course of a week. 

Furthermore, another serious issue is charge compensation, which may originate from intrinsic or extrinsic defects and can be critical for electronic transport and device performance. We also found that atomic force microscopy measurements of MnPS$_3$ \cite{lee2016tunneling} and NiPS$_3$ \cite{kuo2016exfoliation} sometimes show a greater height for the monolayer sample ($\sim 1.5$ nm) than the known interlayer spacing ($\sim 0.6$ nm). This is clearly due to the accumulation of water and oxygen adsorbates at the flake-substrate interface. These intrinsic and surface-generated defects are found to cause trouble in achieving unipolar charge carrier behavior and strong optical exciton signatures in early thin-flake studies. To avoid surface degradation and/or contamination, we always handle all exfoliated flakes of the $TM$PS$_3$ family in an inert atmosphere, preferably in a glovebox. 

\begin{figure*}[t]
\centering
\includegraphics[width=0.8\linewidth]{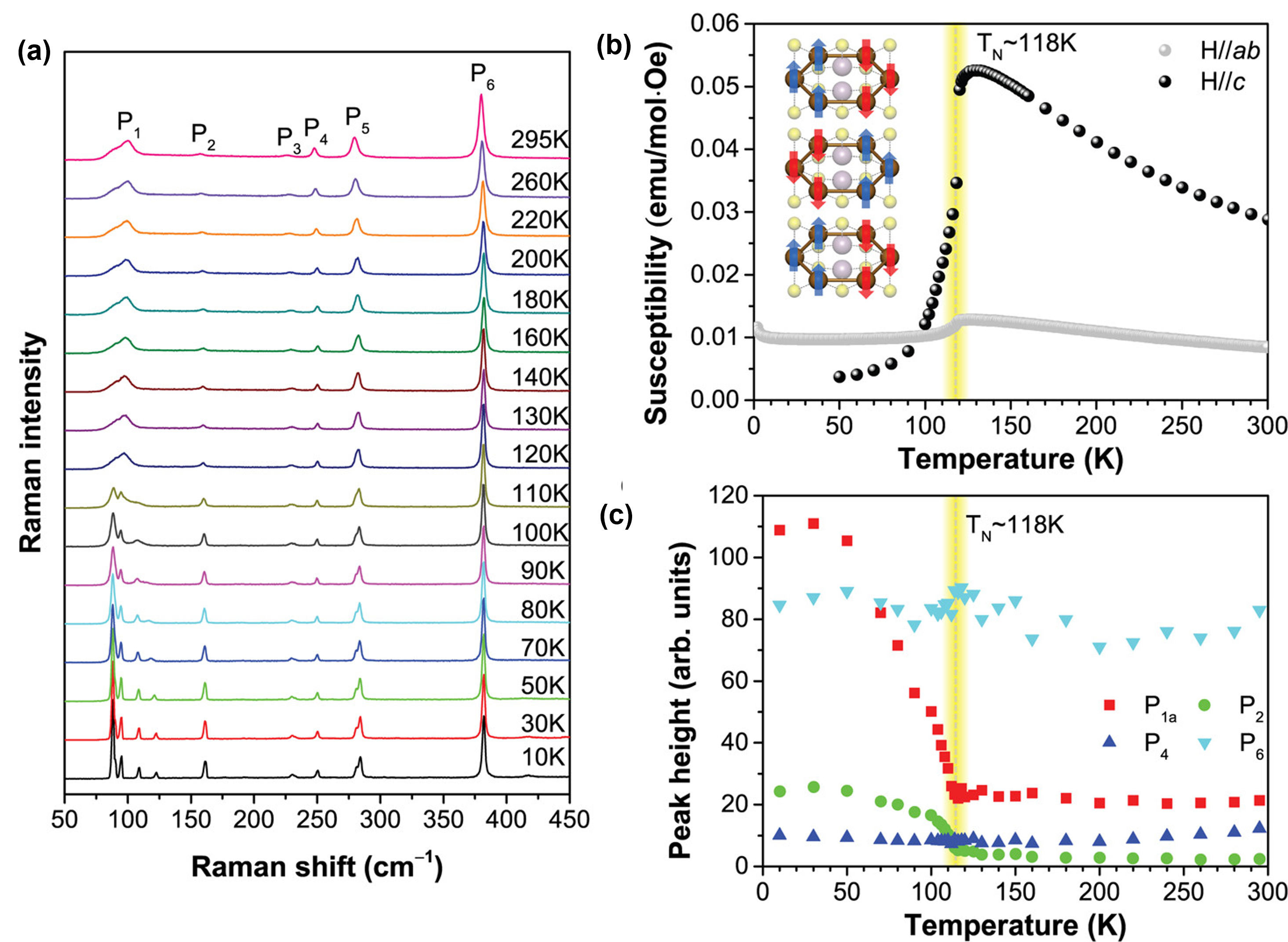}
\caption{Temperature-dependent Raman spectroscopy and magnetic characterization of bulk FePS$_3$. (a) The temperature dependence of the Raman spectrum from 10~K to 295~K reveals six phonon modes; low-frequency modes P$_1$ and P$_2$ (iron vibrations) exhibit pronounced intensity changes across the magnetic transition, whereas higher-energy modes remain relatively stable. (b) Magnetic susceptibility measured along the $ab$-plane (black spheres) and $c$-axis (gray spheres) exhibits a sharp transition at $T_N \approx 118$~K, with the inset showing zig-zag antiferromagnetic alignment of Fe moments. (c) Low-frequency Raman modes (P$_{1a}$ and P$_2$) exhibit dramatic intensity increase below $T_N$. In contrast, higher-energy modes remain temperature-independent, confirming that magnetic ordering induces Brillouin zone folding and couples magnetic and vibrational excitations. The figure is adapted from \cite{lee2016ising}.}\label{fig4}
\end{figure*}

\subsection{Raman Spectroscopy}
Raman spectroscopy has provided unprecedented access to magnetism in 2D materials beyond thickness determination \cite{ferrari2013raman,zhang2015phonon} because of its ability to detect quasi-particles arising from changes in the spin structure. The non-invasive nature and surface sensitivity of Raman spectroscopy make it particularly powerful for probing magnetic systems. This capability is especially important for antiferromagnets, which possess a net-zero magnetic moment, thereby rendering alternative techniques, such as those based on the magneto-optic Kerr effect, ineffective. Another interesting point, not well-appreciated before, is the fact that net long-range magnetic ordering (ferromagnetic or antiferromagnetic) scatters incident light inelastically, as established in the pioneering theoretical work of Fleury and Loudon \cite{fleury1968scattering}. These magnetic excitations can modulate phonon modes, which Raman measurements readily detect. A representative case of probing thickness-dependent magnetism in FePS$_3$, which is one of the first vdW magnets studied in the truly 2D limit, is detailed in the following.

Temperature-dependent polarized Raman measurements were performed down to 10~K on FePS$_3$ samples of varying thicknesses. To identify the magnetic transition temperature, the fundamental principle of Raman spectroscopy was exploited: it can probe energy excitations arising from symmetry breaking. Upon magnetic ordering, rotational symmetry is broken, leading to the emergence of magnons, the quanta of spin-wave excitations. The room-temperature Raman spectra of FePS$_3$ reveal six phonon modes, of which the low-frequency modes P$_1$ and P$_2$ primarily comprise iron vibrations \cite{bernasconi1988lattice}. Below the magnetic transition temperature, spin ordering induces pronounced changes in these modes, as depicted in Fig.~\ref{fig4}(a). This change arises from Brillouin zone folding resulting from zig-zag antiferromagnetic ordering, which maps the $M$ point (from the paramagnetic state) onto the $\Gamma$ point, thus making zone-boundary phonons Raman-active \cite{scagliotti1985spin,scagliotti1987raman,balkanski1987magnetic}. This spectroscopic signature correlates excellently with the magnetic transition observed in the susceptibility data (Fig.~\ref{fig4}(b)), which further indicates Ising-type anisotropy. However, the peak intensities of the higher-energy modes remain unchanged, as shown in Fig.~\ref{fig4}(c), reflecting predominantly molecular-like vibrations from the $(P_2S_6)^{4-}$ bipyramidal structures. Although Raman spectroscopy constitutes a powerful tool for probing magnetism through magnon modes \cite{jin2018raman} and spin-lattice coupling \cite{tian2016magneto}, it provides only qualitative information and cannot quantitatively determine the magnetic order parameter or elucidate the underlying spin structure.

\subsection{Second Harmonic Generation}
Another optical tool that has proven particularly useful for vdW antiferromagnets is second-harmonic generation (SHG). SHG relies on a nonlinear optical process and has been widely used to detect broken inversion symmetry in a given system. For example, when a system undergoes a ferroelectric transition, the SHG is a useful tool to determine the broken inversion symmetry. Likewise, it can be an ideal tool for measuring a similar inversion-breaking effect due to magnetic transitions.

MnPS$_3$ undergoes an antiferromagnetic transition at 78 K, and the type I N\'eel phase breaks the inversion symmetry, unlike FePS$_3$ and NiPS$_3$ \cite{chu2020linear}. The pronounced SHG signal observed in MnPS$_3$ is in stark contrast to that obtained from Raman studies of the same material, which exhibit certain hints of magnetic ordering. However, it is not easy to connect the observed anomaly of the Raman data directly to the magnetic ordering \cite{kim2019antiferromagnetic}. Another notable case is the study of NiI$_2$, the first vdW multiferroic system \cite{ju2021possible,song2022evidence}. In both cases, SHG has been crucial for determining the magnetic phase transition and, consequently, the multiferroic properties down to the bilayer and monolayer, respectively. 

\subsection{Optical Linear dichroism}

Although Raman spectroscopy and SHG measurements can identify magnetic phase transitions and symmetry breaking, they do not have the spatial resolution needed to map the antiferromagnetic domains. Optical linear dichroism, which measures the differential absorption or reflectance of two orthogonally polarized lights, addresses this limitation by probing the electronic anisotropy induced by the magnetic order. Unlike the linear magneto-optical effect, which scales linearly with magnetization, linear dichroism scales with the square of the magnetic order parameter, making it more effective for antiferromagnets.

FePS$_3$ is one such example that exhibits giant optical linear dichroism due to electronic anisotropy linked to the orientation of zigzag chains. Further, angle-dependent linear dichroism measurements can distinguish the three possible zigzag directions \cite{zhang2021observation}. Similarly, linear dichroism has been employed for CrCl$_3$, which is an in-plane antiferromagnet, to determine the in-plane Neel vector orientation and spin-flop transitions under the applied magnetic fields \cite{li2025resolving}. 

\subsection{Advances in magnetic probes }

Although these optical tools are quite handy, one needs local probes to further explore the microscopic physics of these vdW magnets. One requirement is that these probes have a spatial resolution smaller than the characteristic length scale of observables such as magnetic domains and non-collinear spin orders, thereby ensuring accurate resolution of intrinsic magnetic properties. Therefore, a careful selection of the magnetic probe is essential as the specific physical quantities being measured often render certain techniques more suitable for particular systems. 

\begin{figure*}[t]
\centering
\includegraphics[width=0.8\linewidth]{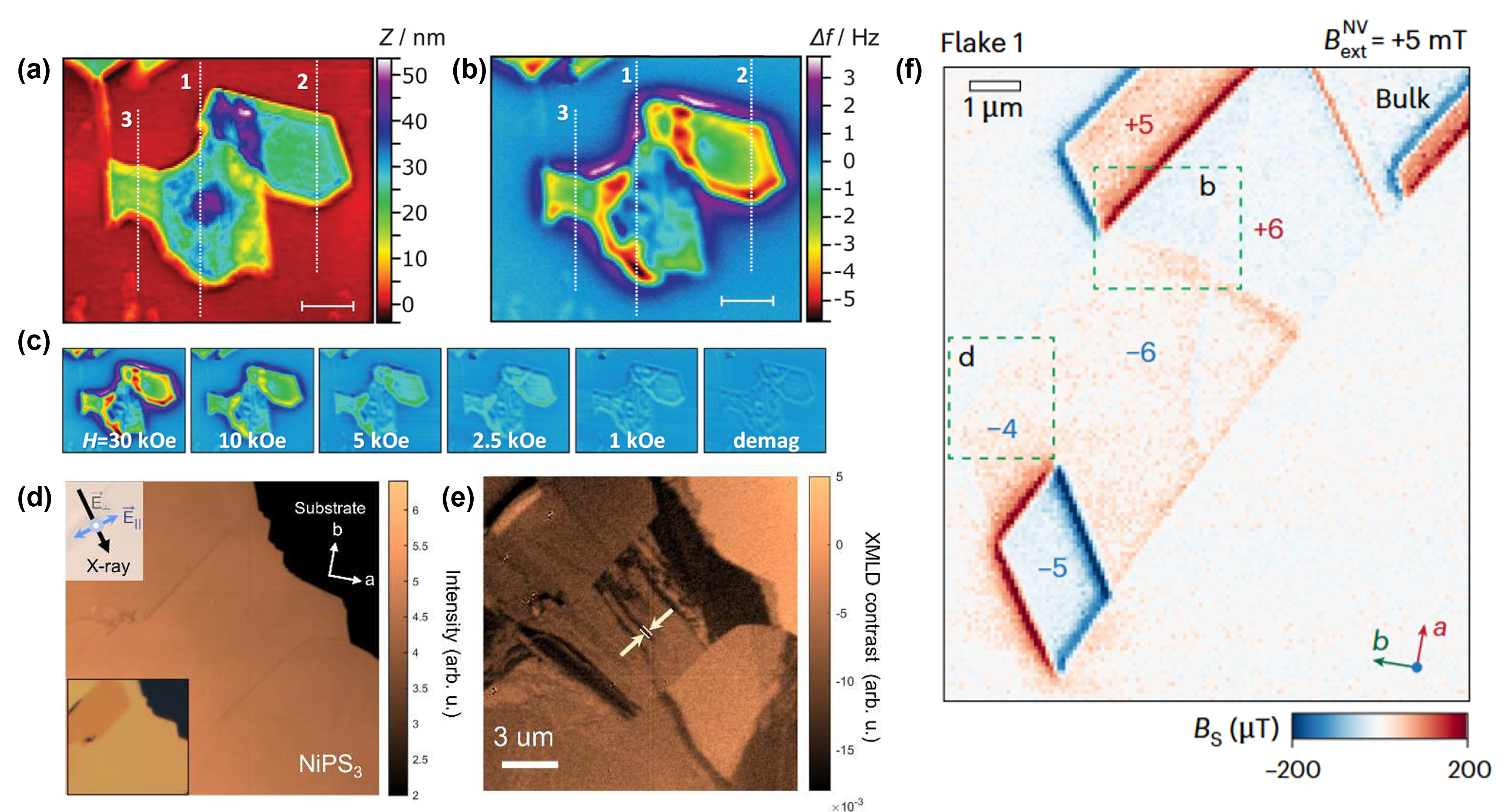}
\caption{(a) Atomic force microscopy (AFM) image of exfoliated CrCl$_3$ flakes on silicon substrate with numbered profiles (1, 2, 3) indicating regions of quantitative analysis. (b) Corresponding magnetic force microscopy (MFM) image acquired at 50~kOe and 14~K, revealing clear magnetic contrast that correlates with flake thickness; dotted lines denote the same profiles as in (a) for direct comparison. (c) Sequence of MFM images acquired at 14~K under progressively decreasing magnetic fields (30~kOe, 10~kOe, 5~kOe, 2.5~kOe, 1~kOe, and demagnetized state), demonstrating field-dependent magnetization reversal. (d) Normalized X-ray photoemission electron microscopy (X-PEEM) image of a 48~nm thick NiPS$_3$ flake recorded at 70~K with photon energy 850.2~eV; the upper inset shows the X-ray incidence direction and electric field polarization geometry, while the lower inset displays the corresponding optical microscope image. (e) X-ray magnetic linear dichroism (XMLD) asymmetry map of the same region, revealing the spatial distribution of magnetic anisotropy and spin orientation. (f) Stray magnetic field ($B_S$) image mapping the nanoscale spin texture; the top and bottom five-layer regions exhibit opposite magnetization states ($|+5\rangle$ and $|-5\rangle$), while the intervening even-layer region displays a weak field signature indicative of antiphase domain walls between $|+6\rangle = \uparrow\downarrow\uparrow\downarrow\uparrow\downarrow$ and $|-6\rangle = \downarrow\uparrow\downarrow\uparrow\downarrow\uparrow$ configurations. Panel (a)-(c) are adapted from \cite{serri2020enhancement}, Copyright \textcopyright 2020 WILEY-VCH Verlag GmbH \& Co. KGaA, Weinheim, panels (d) and (e) are from \cite{lee2024imaging}, and panel (f) is from \cite{wang2025configurable}, Copyright \textcopyright 2025 Nature publications.}\label{fig5}
\end{figure*}

Magnetic force microscopy (MFM), introduced in the 1980s as an extension of atomic force microscopy, represents one such technique. The MFM primarily detects the Lorentz force between a ferromagnetic-coated cantilever tip and the stray magnetic field emanating from the sample. MFM cantilevers are typically manufactured with spring constants comparable to those of atomic bonds, resulting in a high spatial resolution ($\sim 50$ nm) \cite{schmid2010exchange,moser2005high}. However, the mechanical resonance frequency of the cantilever imposes a fundamental upper limit on measurable spin dynamics, restricting the bandwidth to approximately tens of hertz \cite{marchiori2022nanoscale}. Serri \textit{et al.} \cite{serri2020enhancement} demonstrated MFM's applicability to the vdW antiferromagnet CrCl$_3$, observing clear magnetic contrast in flakes with thicknesses ranging from 10 to 50 nm at 14 K under an applied field of 50 kOe, as indicated in Fig.~\ref{fig5}(a) and (b). The magnetic signal faded with decreasing field strength, and analysis of contrast profiles revealed that saturation magnetization remains independent of layer number within this thickness regime, as can be clearly seen in Fig.~\ref{fig5}(c).

X-ray magnetic linear dichroism (XMLD) is a complementary experimental tool. When combined with X-ray photoemission electron microscopy (X-PEEM), it provides atomic-scale monolayer sensitivity and nanometer spatial resolution \cite{lee2023giant,farhan2013exploring,fraile2010size,nolting2000direct}. XMLD exploits the differential absorption between orthogonal linear polarizations (e.g., horizontal versus vertical) in magnetic systems, making it particularly powerful for detecting antiferromagnetic order, non-collinear spin structures, and magnetic anisotropy despite zero net magnetization. XMLD-PEEM measurements have recently been used to image thermally fluctuating N\'eel vectors in NiPS$_3$ \cite{lee2024imaging}. The measurements performed at 70~K on a 48~nm thick flake successfully mapped stripe-like magnetic domains with characteristic dimensions of approximately 200~nm, providing direct access to the N\'eel vector orientation, as exhibited in Fig. \ref{fig5}(e) corresponding to the optical image presented in Fig. \ref{fig5}(d) \cite{lee2024imaging}.

However, scanning nitrogen-vacancy (NV) center microscopy (SNVM) has recently gained substantial interest for imaging various van der Waals magnets. This technique leverages diamond NV centers, which function as single-spin scanning sensors \cite{gruber1997scanning}, with magnetic measurements carried out by optically detected magnetic resonance (ODMR) spectroscopy \cite{schirhagl2014nitrogen}. In this approach, the electron paramagnetic resonance spectrum of the defect center is recorded while the probe is maintained in proximity to the sample surface. SNVM achieves a spatial resolution of approximately 20 nm \cite{ariyaratne2018nanoscale,chang2017nanoscale} combined with an impressive magnetic sensitivity of $\sim 100$ nT\,Hz$^{-1/2}$ \cite{vool2021imaging}, which makes it ideal for probing domain structure and quantifying magnetic strength in vdW antiferromagnets. Furthermore, SNVM can also perform spin relaxometry in the NV resonance region (GHz frequencies) \cite{wolfe2014off,van2015nanometre,bertelli2020magnetic}, which can be extended to kHz-MHz ranges using dynamical decoupling sequences \cite{alvarez2011measuring,degen2017quantum}. Wang \textit{et al.} \cite{wang2025configurable} employed SNVM to reveal antiferromagnetic and ferromagnetic-like regions in CrPS$_4$, corresponding to even and odd layer configurations (see Fig. \ref{fig5}(f)). Additionally, they demonstrated nanoscale magnetization switching, thereby establishing SNVM as a powerful magnetic characterization tool.

\section{Early developments: talks and papers}
\subsection{Two public presentations}

The question of whether magnetism could be stabilised in a true vdW monolayer was notably highlighted by our group at the Korean Physical Society (KPS) meeting in 2015 \cite{park2015kps}. Over the past many decades, the Mermin--Wagner theorem \cite{mermin1966absence} has been a primary source of skepticism regarding experimental attempts to observe 2D magnetism, as it was rigorously established as a fundamental theorem in the late 1960s that long-range magnetic order is forbidden in 2D systems with continuous symmetries. However, Onsager's seminal solution of the 2D Ising model \cite{onsager1944crystal} showed that discrete symmetries and pronounced magnetic anisotropy can circumvent this constraint, enabling stable magnetic ordering at finite temperatures. This theoretical insight suggested that 2D magnets with Ising-type anisotropy might evade the apparent prohibition against ordered states in reduced dimensions. The rise of 2D materials after the discovery of graphene has made the question impossible to ignore. Although the question appeared simple and interesting, it was not easy for one to foresee that its answer could open a new frontier in 2D physics and substantially advance spintronics and quantum device engineering. 

The presentation at the KPS fall meeting in 2015 caused a stir among those who attended, although it was still localized at the time. Subsequent presentations at the Japan--Korea--Taiwan Workshop \cite{park2016jkt} brought attention to a wider community, sharing both intellectual challenges and motivations. This concerted effort strengthened the position of transition-metal phosphorus trisulfides, $TM$PS$_3$ ($TM$~=~Mn, Fe, Co, Ni), as a promising platform for realizing 2D magnetism. These official presentations conveyed to the community that, with a suitable experimental framework, the still-nascent field of 2D vdW magnets can be advanced through rigorous studies.

\subsection{Four Papers}

\begin{figure*}[t]
\centering
\includegraphics[width=0.8\linewidth]{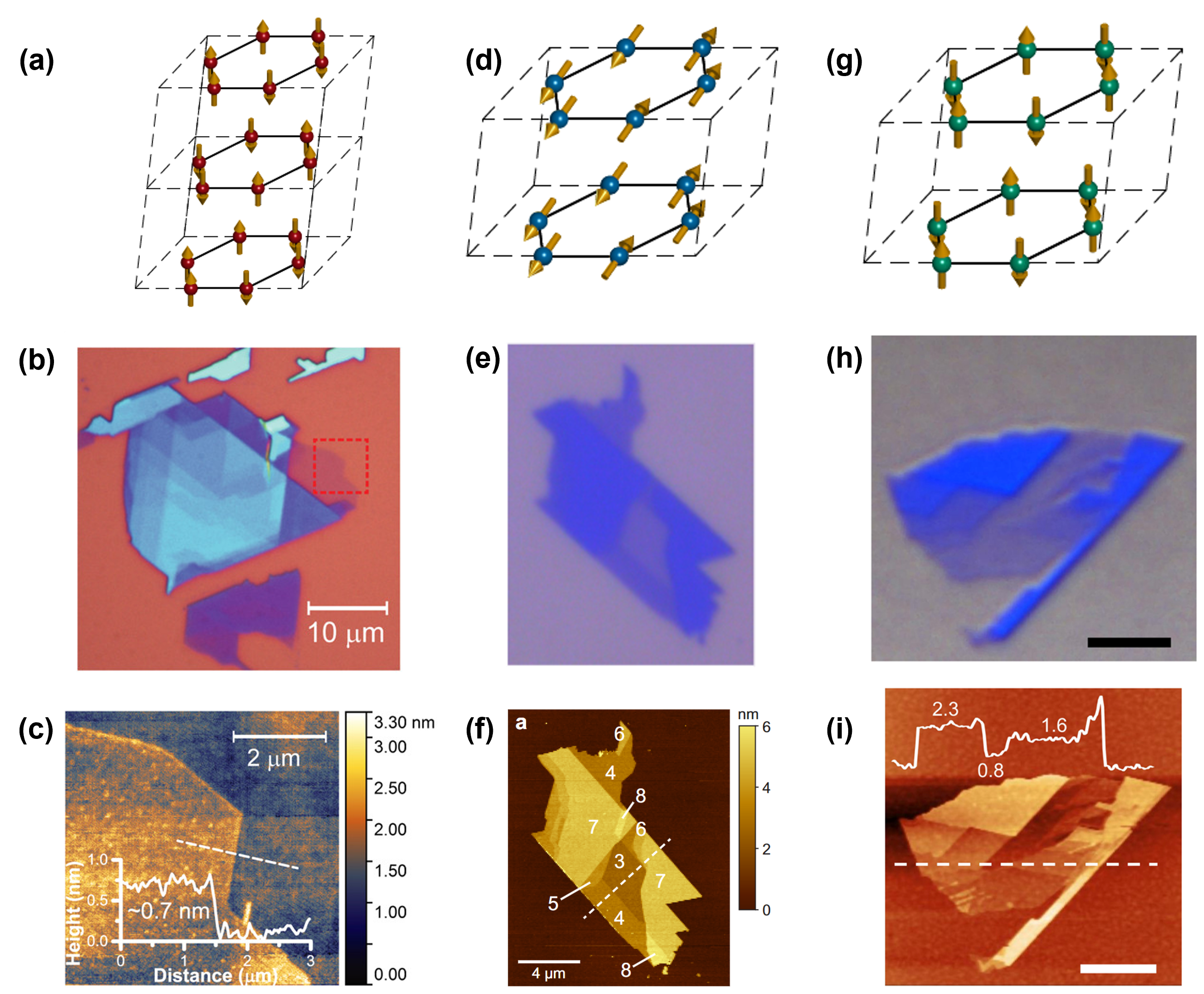}
\caption{(a) Ising-type antiferromagnetic ground state of FePS$_3$. (b--c) Optical contrast image and atomic force microscopy topography of monolayer FePS$_3$ on SiO$_2$/Si substrate, confirming successful mechanical exfoliation to the single-layer limit. (d) XY-type antiferromagnetic configuration of NiPS$_3$. (e--f) Brightfield optical microscopy and corresponding tapping-mode AFM topography of thin NiPS$_3$ sheets with layer numbers (3--8 layers) clearly identified from height analysis. (g) Heisenberg-type antiferromagnetic spin structure of MnPS$_3$. (h--i) Optical microscopic and AFM topographic images of MnPS$_3$ flakes ranging from monolayer to hexalayer, with quantitative height profiling revealing the progressive increase in thickness and corresponding structural features. Panel (a),(d) and (g) are adapted from \cite{park20252d}; panels (b--c) are from \cite{lee2016ising}; panels (e--f) are from \cite{kuo2016exfoliation} and panels (h--i) are from \cite{lee2016tunneling}.}\label{fig6}
\end{figure*}

In 2016, four key early papers from our group established exfoliation protocols and provided the first monolayer-scale evidence for antiferromagnetic order.  In a work of NiPS$_3$ \cite{kuo2016exfoliation}, Kuo \textit{et al.} demonstrated the first successful mechanical exfoliation of the vdW magnet, NiPS$_3$, to the monolayer limit using the conventional scotch-tape methodology. The authors employed optical contrast, Raman spectroscopy, and atomic force microscopy to identify flakes and determine the number of layers with precision. In particular, tracking the out-of-plane sulfur mode, $A_{1g}^{(1)}$, provided qualitative layer identification. Although no direct magnetic measurements were reported, this work established $TM$PS$_3$ as a versatile platform for 2D magnetism. The chemical flexibility of this family, as mentioned above, combined with the possibility of substituting sulfur with selenium, has created an rich playground for exploring 2D magnetic phenomena.

Building on this success, our group subsequently investigated MnPS$_3$, another $TM$PS$_3$ family member that is amenable to monolayer exfoliation, employing conductive atomic force microscopy (C-AFM) for transport characterization \cite{lee2016tunneling}. Tunneling transport measurements were conducted on MnPS$_3$ flakes ranging from monolayers to tens of layers deposited on a 70~nm indium tin oxide (ITO) layer on a silicon substrate. These measurements conclusively confirmed the insulating character of MnPS$_3$ in all thicknesses down to the monolayer limit, thereby extending the scope of magnetic tunnel junction research to encompass 2D magnetic materials.

In the same year, our group also turned its attention to FePS$_3$, conducting comprehensive magnetism measurements in the truly 2D limit using Raman spectroscopy as the primary probe \cite{lee2016ising}. As detailed in the preceding Raman spectroscopy subsection, Raman measurements unambiguously confirmed long-range magnetic ordering that persists down to the monolayer limit in FePS$_3$. This result provided experimental verification of Onsager's solution \cite{onsager1944crystal} for 2D order-disorder transitions: while the Mermin--Wagner theorem \cite{mermin1966absence} forbids long-range ordering in XY and Heisenberg models at finite temperatures in reduced dimensions, Onsager demonstrated that 2D Ising systems can sustain long-range magnetic order at finite temperatures. This work thus constituted the first atomically thin, exfoliable vdW magnet (monolayer). Figure \ref{fig6} presents a unified summary of the three seminal studies, showing the magnetic structures and the minimum exfoliated thickness.

During the discoveries of antiferromagnetism, ferromagnetic ground states were reported in Cr$_2$Ge$_2$Te$_6$ \cite{gong2017discovery} and CrI$_3$ \cite{huang2017layer} in 2017, establishing that 2D vdW magnets could exhibit both conventional long-range order. The demonstration of both magnetic states caused a global research surge in 2D magnetism, rapidly expanding the field to encompass novel phenomena such as spin-filtering in CrI$_3$ tunnel barriers \cite{klein2018probing,song2018giant}, gate-tunable magnetism \cite{deng2018gate}, and the emergence of 2D magneto-optical effects. International participation expanded dramatically across multiple research frontiers, and heterostructure engineering (graphene, transition-metal dichalcogenides, moir\'e superlattices) has emerged as one of the most prominent directions \cite{tong2018skyrmions,hejazi2020noncollinear,xu2022coexisting}.

This explosive growth in the field was articulated in the seminal viewpoint by Park \cite{park2016opportunities}, who articulated the opportunities and challenges of 2D magnetic vdW materials, posing the question: ``magnetic graphene?'' This perspective further recognised the transformative potential of 2D magnetism for fundamental physics and technological applications, solidifying vdW magnets as a central pillar of contemporary condensed-matter research.

\section{New physics enabled by 2D magnets}

\subsection{Dimensional crossover and magnetic criticality}

In the 2D limit, there are three principal mechanisms of magnetic ordering. They are (1) magnetic anisotropy, (2) dipolar interactions, and (3) competing exchange pathways. First, magnetic anisotropy suppresses the logarithmic divergence of thermal fluctuations predicted by the Mermin--Wagner theorem \cite{mermin1966absence}. Second, dipolar interactions, although weak, stabilize long-range order and reinforce anisotropy effects. Finally, competing exchange pathways enable tunable control over magnetic properties.

Magnetic anisotropy arises from the interplay between crystal-field splitting and spin-orbit coupling, thereby establishing spin structures with preferred orientations. For example, FePS$_3$ exhibits large single-ion anisotropy ($\sim 20$ meV), which stabilizes Ising-type antiferromagnetic order with an easy out-of-plane axis \cite{lee2023giant}. This dimensional reduction of spin space from three to one profoundly suppresses thermal fluctuations. In contrast, the relatively weak single-ion anisotropy in NiPS$_3$ \cite{kim2019suppression} and CoPS$_3$ \cite{wildes2017magnetic} leads to XY-type antiferromagnetic order, confining spins to a single plane.

Dipolar interactions have been proposed as a key stabilizing mechanism for the long-range antiferromagnetic order in NiPS$_3$ monolayers. Theoretical studies reveal a complex exchange interaction hierarchy \cite{kim2021magnetic}: the nearest-neighbor coupling is ferromagnetic ($\sim -2.5$ meV), the second nearest-neighbor interaction is weakly antiferromagnetic ($\sim 0.2$ meV), and the third nearest-neighbor coupling is strongly antiferromagnetic ($\sim 13.9$ meV), with the latter dominating the ground state \cite{scheie2023spin}. This exchange frustration suggests that dipolar interactions having magnitudes comparable to exchange parameters may provide essential long-range coherence and anisotropic stabilization that complement single-ion anisotropy in the monolayer limit \cite{kim2021magnetic}.

The exchange pathways further illuminate the role of the bond geometry and electronic configuration in stabilizing magnetic order. For example, MnPS$_3$ magnetism is dominated by nearest-neighbor interactions originating from $t_{2g}$ orbital magnetism \cite{autieri2022limited,wildes1998spin,olsen2021magnetic}. By contrast, NiPS$_3$ exhibits third nearest-neighbor exchange as the strongest interaction, a consequence of the filled $t_{2g}$ subshell and strong hybridization between the $d_{x^2-y^2}$ orbitals and sulfur $p$-orbitals within the same atomic layer \cite{autieri2022limited}. These distinct exchange hierarchies yield fundamentally different ground states, the N\'eel order in MnPS$_3$ versus the zig-zag order in NiPS$_3$, accompanied by widely separated magnetic transition temperatures (78~K versus 155~K), demonstrating how subtle variations in electronic structure encode dramatic consequences for 2D magnetism.  

\subsection{Optical control of magnetic anisotropy}

\begin{figure*}
\centering
\includegraphics[width=0.8\linewidth]{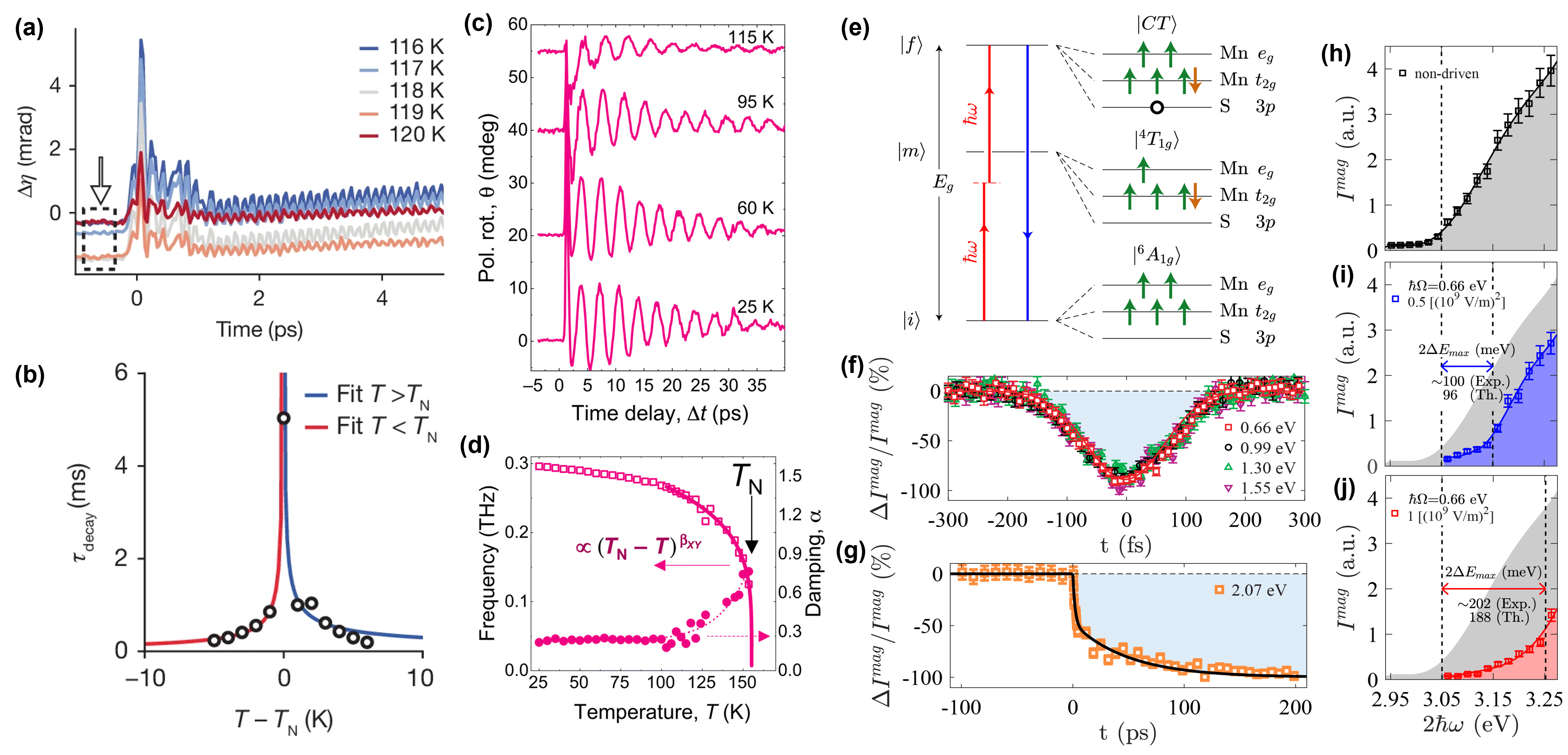}
\caption{(a--b) FePS$_3$: Pre-time-zero ellipticity signal near the N\'{e}el temperature reveals metastable-state formation; decay-time analysis demonstrates critical exponent behavior ($z\nu = 0.89$ above $T_N$ and $0.56$ below $T_N$), indicative of magnetic criticality. (c--d) NiPS$_3$: Temperature-dependent time-resolved polarization rotation following 1.08~eV pump excitation exhibits pronounced oscillations; the extracted magnon frequency follows a power law $\propto (T_N - T)^{b\kappa\nu}$, demonstrating magnon softening at the magnetic transition. (e) Schematic energy diagram for resonant electric-dipole second-harmonic generation (ED-SHG) in MnPS$_3$: multi-electron states (left) and corresponding orbital/spin configurations (right) illustrate the hybridization of ground ($^6A_{1g}$) and charge-transfer states ($|CT\rangle$) under optical driving. (f) Transient SHG response ($\Delta I_{\text{mag}}/I_{\text{mag}}$) for subgap pump photon energies ($\hbar\omega = 0.66$--$1.55$ eV) at fixed electric-field amplitude ($E_{\text{max}} = 10^9$ V/m). (g) SHG transient under resonant pumping conditions at photon energy 2.07~eV. (h--j) Non-driven (h) and pump-driven SHG spectra at $t=0$ for increasing field amplitudes (i) $(E_{\text{max}})^2 = 0.5 \times 10^{18}$ V$^2$/m$^2$ and (j) $1.0 \times 10^{18}$ V$^2$/m$^2$, demonstrating field-dependent modification of the anisotropic response and validating the Floquet mechanism for dynamic control of spin-orbit admixture. Panels (a--b) are adapted from \cite{ilyas2024terahertz}, panels (c--d) are from \cite{afanasiev2021controlling} and panels (e--j) are adapted from \cite{shan2021giant}.}\label{fig7}
\end{figure*}

In 2D magnets, magnetic anisotropy not only dictates spin ordering geometry but also protects magnetic order from thermal fluctuations, which intensify with dimensional reduction. This intricate coupling between magnetic order and anisotropy motivates the development of strategies to dynamically control anisotropy, thereby providing unprecedented access to manipulate magnetic ordering. As described in section \ref{sec2}, the strength of magnetic anisotropy is generally determined by the coupling between the electronic orbital angular momentum and spin. However, most 2D magnets based on transition-metal ions possess partially or fully quenched orbital angular momentum \cite{afanasiev2021controlling}. One powerful approach to controlling magnetic anisotropy is optical pumping to higher-energy orbital states \cite{mikhaylovskiy2015terahertz,baierl2016nonlinear,iida2011spectral,mikhaylovskiy2020resonant}. For example, FePS$_3$ exhibits an unquenched orbital moment ($L \sim 1$), allowing anisotropy control through optical resonance of low-energy magnon and phonon modes to higher amplitudes. This nonlinear driving of specific phonon modes modifies in-plane distances between magnetic ions, establishing a metastable state (see Fig. \ref{fig7}(a--b)) with an impressive lifetime of 2.5~milliseconds \cite{ilyas2024terahertz}.

On the other hand, in systems with zero or negligible unquenched orbital angular momentum, magnetic anisotropy arises from spin-orbit-driven mixing of the ground state with higher-energy states bearing residual orbital character. For example, NiPS$_3$ has a quenched orbital moment ($L = 0$), and magnetic anisotropy emerges from spin-orbit-mediated hybridization of the ground state ($^3A_{2g}$) with the first excited state ($^3T_{2g}$) \cite{joy1992magnetism,chandrasekharan1994magnetism}. In such cases, magnetic anisotropy can be controlled by resonating photon energy with orbital transitions among crystal-field-split levels, as depicted in Fig. \ref{fig7}(c--d). These magnon modes, activated by subterahertz frequencies, are used to perform polarization-dependent control of the magnon amplitude, confirming the dynamic manipulation of magnetic anisotropy \cite{afanasiev2021controlling}.

The weak magnetic anisotropy ($\sim 0.009$ meV) of MnPS$_3$ is due to the mixing of the ground state ($^6A_{1g}$) with excited states bearing a residual orbital character. A recent control of magnetic anisotropy in this system employs a Floquet method \cite{basov2017towards,oka2019floquet}, which bypasses the issue of heating by using subgap wavelengths. Shan \textit{et al.} \cite{shan2021giant} showed that optical pumping of MnPS$_3$ at subgap frequencies ($\hbar\omega = 0.66$--$0.99$ eV) resonant with crystal-field transitions hybridizes the ground state and charge-transfer states through a time-dependent oscillating field ($\sim 10^9$ V/m), (see Fig. \ref{fig7}(g--j)). This dynamic modulation of the crystal field modulates the spin-orbit admixture coefficients, allowing real-time control of both the strength and orientation of the magnetic anisotropy measured by SHG.  

\subsection{Moir\'e magnetism: Stacking angle engineering}

An alternative pathway to unlock tunable magnetism in vdW magnets exploits stacking-angle engineering through moir\'e superlattices created by rotating or translating individual 2D magnetic layers relative to each other. In particular, small stacking angles can result in various stacking geometries that cannot be realized in a pristine sample. The presence of these high-symmetry configurations with distinct interlayer exchange interactions provides an additional control parameter \cite{hejazi2020noncollinear,tong2018skyrmions,wang2020stacking}. The bilayer CrI$_3$ is an archetypal antiferromagnet for such stacking engineering in which exchange interactions and the resulting magnetic state depend on the type of stacking, whether it is monoclinic (AB$'$) or rhombohedral (AB) stacking \cite{song2021direct}. This results in a noncollinear spin structure that exhibits in-plane spin canting, with magnetic order perpendicular to the plane, which can be further classified as bubble-like excitations or magnetic skyrmions.

\subsection{Magnetic heterostructures}

One notable trend in recent work on vdW magnets is the study of their heterostructures with 2D conductors and semiconductors. A large volume of recent work has focused on proximity effects, aiming to realize new functional capabilities for future device applications \cite{zhang20242d}. A clear strategy has been the use of heavy metals, such as tungsten (W) or platinum (Pt), which possess strong spin-orbit coupling \cite{su2020current,ye2022nonreciprocal,lohmann2019probing}. In this configuration, the heavy metals can be used to enhance the effectiveness of magnetization detection probes for antiferromagnets. If successful, it could circumvent the challenges posed by zero net magnetization and the inherent limitations of conventional detection techniques. In this regard, it should be noted that Ma \textit{et al.} demonstrated substantial spin polarization in the Pt layer of FePS$_3$ / Pt heterostructures, resulting in a measurable anomalous Hall effect \cite{ma2025van}. Furthermore, the temperature dependence of the anomalous Hall response reproduces the magnetic phase transition of FePS$_3$, providing elegant proof of proximity-induced magnetization transfer.

A particularly promising recent development involves the manipulation of the topological state using ionic gating, as reported for Co$_{1/3}$TaS$_2$, an air-stable intercalated non-coplanar antiferromagnet with intrinsic topological spin texture \cite{kim2025electrical}. Zhang \textit{et} \textit{al.} \cite{zhang2025current} further demonstrated control of topological spin chirality in Co$_{1/3}$TaS$_2$ by integrating it with Pt, which leads to spin-Hall-effect-induced nonvolatile and reversible switching. Interestingly, Co$_{1/3}$TaS$_2$ exhibits an intrinsic self-spin orbit torque arising from broken inversion symmetry and pronounced Berry curvature hotspots, enabling field-free and heavy-metal-free electrical switching of topological spin chirality. This dual-pathway approach, which combines interfacial spin-orbit torque or intrinsic self-spin-orbit torque, establishes new paradigms for topological spintronics and quantum magnetic device engineering.

Exchange bias effects, which emerge at the ferromagnetic/antiferromagnetic interfaces, represent another important avenue for heterostructure engineering and manipulation for spintronics applications \cite{zhu2020exchange,hu2020antisymmetric,dai2021enhancement}. Recently, Albarakati \textit{et al.} demonstrated that the exchange field in FePS$_3$/Fe$_3$GeTe$_2$ heterostructures can be dynamically controlled using solid protonic gating, allowing electrical modulation of interfacial magnetic coupling \cite{albarakati2022electric}. The exchange bias field can also be tuned by changing the thickness of the ferromagnet, as reported for Fe$_{5-x}$GeTe$_2$/CrPS$_4$ heterostructure \cite{wei2025exchange}. Moreover, X. Wang $et$ $al.$ have shown pressure control of emergent interfacial antiferromagnetism in Fe$_3$GeTe$_2$/MnPS$_3$ heterostructure, which goes beyond the conventional effect of exchange bias \cite{wang2025artificially}.

Magnetoresistance, defined as a change in electrical resistance on application of a magnetic field, can also provide key insights into the interfacial physics. For instance, in the case of the MnPS$_3$/Fe$_3$GeTe$_2$ heterostructure, three distinct resistances are observed as a possible consequence of unsynchronized magnetic switching between the antiferromagnetic layer and ferromagnetic interface and bulk of the ferromagnet \cite{hu2020antisymmetric}. A large negative magnetoresistance is reported in CrSBr, a layered antiferromagnet, which can be tuned by gating \cite{telford2020layered,chou2025large}. Moreover, twisted bilayers of CrSBr exhibit a tunneling magnetoresistance exceeding 700 \%, indicating their potential for high-storage spintronic devices \cite{chen2024twist}. Similarly, a nonvolatile but multistate magnetoresistance has also been observed in  (Fe$_{0.8}$Co$_{0.2}$)$_3$GaTe$_2$/WSe$_2$/(Fe$_{0.8}$Co$_{0.2}$)$_3$GaTe$_2$, where (Fe$_{0.8}$Co$_{0.2}$)$_3$GaTe$_2$ is an A-type antiferromagnet with a transition temperature around 210 K \cite{jin2025nonvolatile}. Interestingly, another A-type antiferromagnet, CrPS$_4$, shows oscillatory magnetoresistance which is intrinsic in nature and may be due to a canted magnetic state under applied magnetic fields \cite{shi2024magnetoresistance}.

\section{Toward 2D Spintronic and Quantum Devices} 

Ferromagnetic materials have traditionally been heavily used in spintronics devices. However, the increasing demand for low-power consumption devices, high storage density, and faster read/write speeds has recently caused renewed interest in alternative antiferromagnetic platforms \cite{jungwirth2016antiferromagnetic,han2023coherent,dal2024antiferromagnetic}. In this recent context, antiferromagnets promise several advantages over ferromagnets: for example, ultrafast spin dynamics, robustness against external magnetic fields, and negligible stray-field radiation \cite{yan2020electric,rahman2021recent}. Among antiferromagnets, there are two different classes: one with collinear spin structures and another with non-collinear spin textures. 

A recent example of such collinear antiferromagnets is CrSBr, an A-type collinear antiferromagnet that exhibits intrinsic net-zero spin polarization \cite{yao2025switching,gong2018electrically,liu2024electronic}. Carrier doping is shown to modify the interlayer exchange coupling and stabilize a highly spin-polarized ferromagnetic phase. For example, the magnetic order of CrSBr could be spatially modulated by combining a doping approach with local interfacial charge transfer from graphene contacts. An interesting point is that the resulting lateral FM/AFM/FM configurations are, in principle, gate-tunable spin valves with a complete reversal of spin polarization across a critical gate voltage without requiring ferromagnetic metal contacts \cite{zhao2025doping}. This represents a major simplification with respect to the integrated device architectures of conventional spintronics. However, the inherent unresponsiveness of antiferromagnets to external magnetic perturbations remains a significant technical challenge for controlling and probing magnetic order.

At the same time, non-collinear antiferromagnets offer a potential solution by synergistically combining the advantages of ferromagnets (efficient reading and writing through detectable magnetization) with those of antiferromagnets (ultrafast dynamics, negligible stray fields, and intrinsic non-volatility) while simultaneously circumventing their respective drawbacks. Two such candidates are Ni$_{1/3}$NbS$_2$ and Co$_{1/3}$TaS$_2$. In the case of Ni$_{1/3}$NbS$_2$, the application of an in-plane current leads to the generation of antiferromagnetic spin-orbit torque, which then acts as a controller of the entire helical order \cite{zhang2025current1}. Meanwhile, in the case of Co$_{1/3}$TaS$_2$, controlling its topological 3$Q$ state, which manipulates the associated spontaneous topological Hall effect, can open new pathways for the efficient readout of antiferromagnetic states electrically \cite{takagi2023spontaneous,park2023tetrahedral}. Furthermore, Meng \textit{et al.} have demonstrated successful integration with the dichalcogenide MoS$_2$, revealing its potential for future spintronic devices \cite{meng2025spontaneous}.

\section{Outlook and future prospects}

\begin{figure*}
\centering
\includegraphics[width=0.8\linewidth]{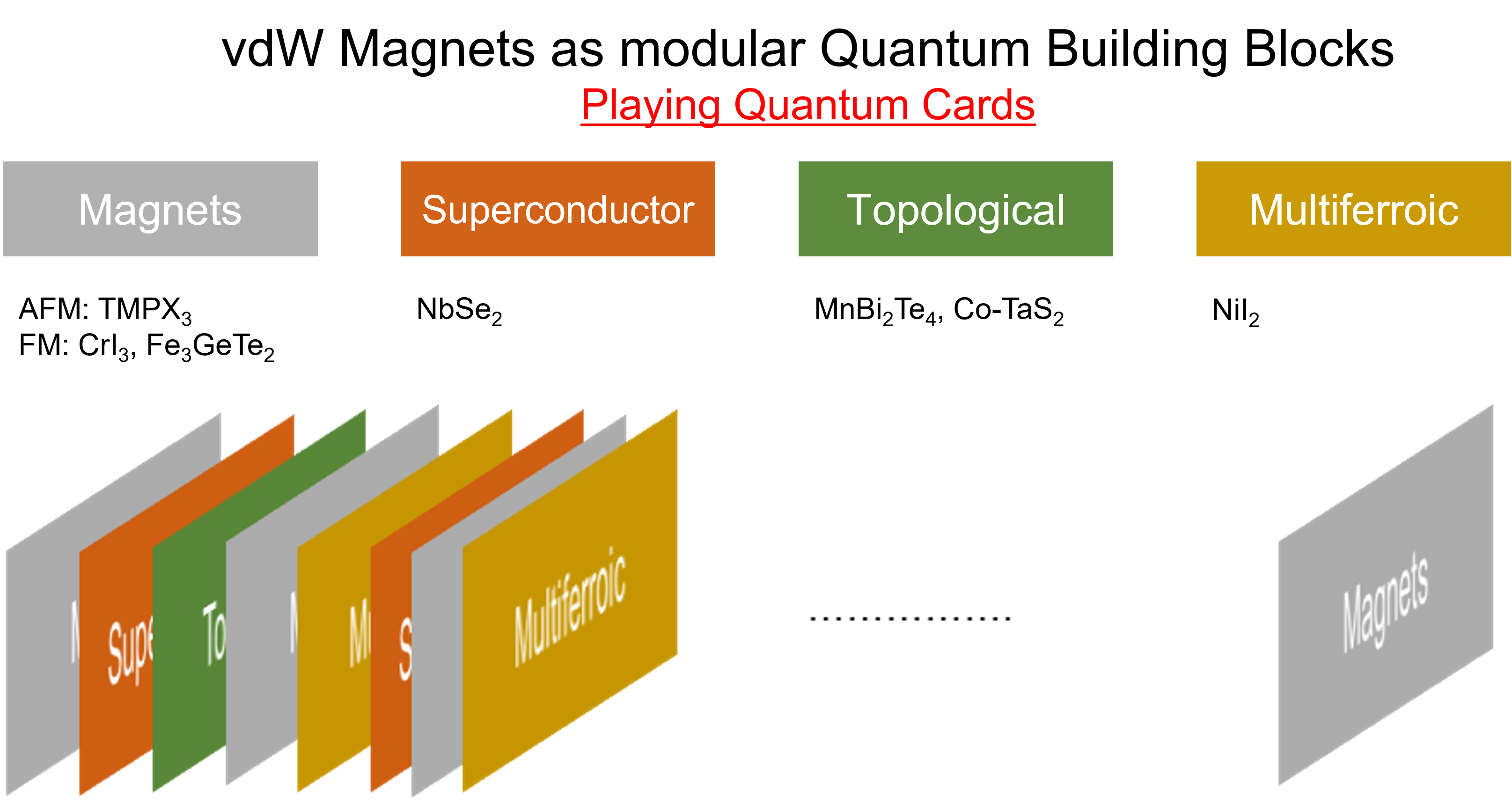}
\caption{As a conceptual organizing scheme for future studies of vdW magnets.}\label{fig8}
\end{figure*}

In the case of 2D materials, vdW magnets will further strengthen their position as building blocks, especially through interface engineering and light-matter interaction, in the upcoming decade. The vdW magnets can be considered as an intersection space of nanoscale condensed matter physics and spintronics. The exploration of this space will advance our existing understanding of nanoscale magnetism and how these could be utilized for next-generation practical devices. As illustrated in Fig. \ref{fig8}, when vdW magnets are combined with properties such as superconductivity, topology, or multiferroicity, referred to as 'quantum cards', individually or in combination, novel emergent phenomena that were previously inaccessible can be studied.

\subsection{Robustness of bulk properties in 2D limit}

First of all, we would like to emphasize an important aspect which is often overlooked:  whether a vdW magnet retains its bulk properties down to the exfoliated limit. Generally, properties in the thin limit remain the same as in bulk, but one should be careful when interpreting results in the 2D limit. For instance, a few-layer CrI$_3$ favours a monoclinic stacking and hence an antiferromagnetic ground state, while it has a ferromagnetic ground state in bulk form owing to the rhombohedral stacking \cite{song2019switching,li2019pressure}. Similarly, one should be careful about the potential substrate effects which may contribute to their magnetic properties. Hence, it is necessary to do structural scrutiny also to confirm that the observed phenomena are intrinsic in nature.

\subsection{Beyond Three Fundamental Spin Models in the 2D Limit}

The discovery of vdW antiferromagnets has enabled the experimental realization of Ising, XY, and Heisenberg spin models with unprecedented tunability and control. These systems are expected to enable the further exploration of exotic spin phases, such as spin-glass \cite{binder1986spin} and spin-liquid states \cite{anderson1987resonating}, in the 2D limit. However, such disordered phases, characterized by quenched disorder, frustration, and the absence of long-range order, remain largely unexplored in vdW systems. A fundamental open question is whether the spin-glass or spin-liquid phases retain stability when reduced to the 2D limit \cite{hoines1991thin}.

In the case of spin-glass states, theoretical studies indicate that, although the 2D random Ising model lacks a finite-temperature transition, it may host a zero-temperature spin-glass transition at or above two dimensions \cite{morgenstern1980magnetic}. For XY and Heisenberg spin glasses with short-range interactions, two dimensions fall below the critical dimensionality required to sustain such a phase. This renders the experimental realization of 2D spin glasses particularly challenging, though not fundamentally impossible.

Quantum spin liquid states present an equally compelling frontier. Recent reports on monolayer 1T-NbSe$_2$ and 1T-TaSe$_2$ suggest signatures of quantum spin liquid behavior \cite{zhang2024quantum,ruan2021evidence}; however, a definitive identification of the ground state remains ambiguous. The search for robust quantum spin-liquid signatures in van der Waals materials will undoubtedly occupy a central position in future research.

\subsection{Strongly correlated physics in magnetic moir\'e superlattices}

The emergence of flat bands in twisted bilayer vdW antiferromagnets represents a promising frontier for engineering strongly correlated magnetic phases with unprecedented tunability and control. This motivation is anchored in the "magic angle" physics reported in twisted bilayer graphene, where a twist angle of $\sim$ $1.1^\circ$ stabilizes flat electronic bands \cite{cao2018correlated}, which in turn enables superconductivity \cite{cao2018unconventional,behura2021moire}. Twisted bilayer antiferromagnets similarly promise engineered magnetic correlations with dynamically tunable interaction strengths. Recent first-principles calculations predict the emergence of flat bands and quantum phase transitions from antiferromagnetic to ferromagnetic states in twisted CrSBr bilayers \cite{liu2025moire}. Furthermore, multiple moir\'e heterostructures can be investigated, whose interference can generate secondary periodicity with length scales substantially larger than either parent moir\'e structure. In fact, it has been observed in a twisted (twist angle = $\sim$ $1.1^\circ$) double bilayer of CrI$_3$, with a length scale of $\sim$300 nm, which spans across multiple parent moir\'e cells \cite{wong2025super}.

\subsection{Non-equilibrium magnetic control}

Creating metastable magnetic states in vdW antiferromagnets and controlling them on timescales of picoseconds to femtoseconds can lead to ultrafast memory devices. One way to access these states is through light-matter coupling; however, future device physics will require a deeper understanding of this interaction, which can be gained through the study of nonlinear optical responses. In particular, recent studies on Co$_{1/3}$TaS$_2$ reveal that the nonlinear optical signals are related to their geometric phase and thus can be used to investigate the topological magnetic properties \cite{bao2022light}.

However, light can heat the sample under study and degrade it. This problem has been solved by Floquet engineering, which has recently gained significant attention and enables the control of a metastable topological state that was previously unreachable under equilibrium conditions. To further extend its capabilities, significant advances in instrumentation are required, particularly in electromagnetic wave sources. Moreover, extending the lifetime of the coherent state will be crucial for manipulating a metastable state for longer \cite{bao2022light}. Hence, technological advances in the Floquet method are necessary to realize its full potential.

\subsection{Integration into vdW devices}

As noted above, the integration of antiferromagnets with other functional materials can enable energy-efficient, fast switching. Together, they will represent a potentially transformative development for spintronics technologies. For example, Mn$_3$Sn, a bulk topological antiferromagnet, reduces the switching current by an order of magnitude compared to ferromagnetic systems when interfaced with platinum of optimized thickness \cite{zheng2024effective}. We anticipate that further optimization, achieved through the careful selection of antiferromagnetic materials, will yield even more substantial improvements. Moreover, Mn$_3$Sn can closely mimic a synapse, indicating its potential to approximate neural networks and the potential of such antiferromagnets for neuromorphic computing applications \cite{wang20202d}.

An important requirement for practical spintronic application is the robust operation of the devices at room temperature. While enhancing the Neel temperature of antiferromagnetic materials remains a challenge, recent efforts to substitute Co for Fe in Fe$_5$GeTe$_2$ offer a promising pathway. The Co substitution in a particular range (40 -- 45\%) results in a structural transition of Fe$_5$GeTe$_2$ from rhombohedral ($R\bar{3}m$) to primitive ($P\bar{3}m1$) space group, which lead to a crossover from ferromagnetic state to antiferromagnetic state with a Neel temperature above room temperature: (Fe$_{0.56}$Co$_{0.44}$)$_5$GeTe$_2$ ($T_N$ $\sim$ 335 K) \cite{may2020tuning,tian2020tunable,zhang2022room} and (Fe$_{0.6}$Co$_{0.4}$)$_5$GeTe$_2$ ($T_N$ = 374 K) \cite{hu2024enhanced}. However, its dimensional scaling reveals that the antiferromagnetic order sustains down to a thickness of four layers, below which the system reverts to a ferromagnetic state \cite{lu2024tunable}. Despite this thickness constraint, W. M. Zhao $et$ $al.$ have shown that the (Fe$_{0.6}$Co$_{0.4}$)$_5$GeTe$_2$/WSe$_2$/(Fe$_{0.6}$Co$_{0.4}$)$_5$GeTe$_2$ heterostructure with 4--6 layers thick flakes could exhibit magnetoresistance up to 230 K, which signals a potential for high-temperature spintronic devices \cite{zhao2025interface}. 

Spin-caloritronic phenomena, which originate from a temperature gradient leading to spin current generation or vice-versa, can also provide important insight into the magnetic properties. For example, a heterostructrue of CrPS$_4$ with Pt reveals the already established spin-flip transition while also demonstrating magnon diffusion above 6~$\mu$m \cite{tian2025thermal,he2025spin}, making it a promising platform for advanced magnonics.

As shown in Fig. \ref{fig8}, integrating the vdW antiferromagnet with different systems can reveal novel physics at the interface through the proximity effect. In particular, interfacing vdW antiferromagnets with a superconductor may lead to the emergence of skyrmion-vortex pairs and to the hosting of a topological superconducting state. It is well established that such a state can have Majorana zero modes, which are the backbone of fault-tolerant quantum computing. One such candidate among the currently available vdW antiferromagnets is Co$_{1/3}$TaS$_2$, which already hosts the spontaneous Hall effect, and its combination with a suitable superconductor can lead to the realization of novel topological states. 

\subsection{Charge Transfer Physics}

Our final word concerns a particular aspect of physics that has been largely ignored in the vdW magnetism community: the effect of small or negative charge transfer. Arguably, most of our knowledge of modern magnetism comes from decades-long research into magnetic oxides, where five key energy scales govern electronic and magnetic properties: the Coulomb repulsion $U$ ($\sim 3$--$8$ eV), Hund's coupling $J_H$ ($\sim 0.8$--$1$ eV), crystal-field splitting $10D_q$ ($\sim 2$--$4$ eV), charge-transfer energy $\Delta$ ($\sim 3$--$5$ eV), and superexchange $J$ ($\sim 100$ meV). In contrast, van der Waals magnets having chalcogenide and halide elements exhibit substantially different scales: $U \sim 3$--$8$ eV, $J_H \sim 0.8$--$1$ eV, $10D_q \sim 1$--$3$ eV, $\Delta \sim -1$--$1$ eV, and $J \sim 10$ meV.

Among these physical parameters, the most crucial is the charge-transfer energy ($\Delta$) \cite{zaanen1985band,khomskii2001unusual,bocquet1996electronic}. The relative ratio of $\Delta$ and $U$ has been used to classify oxides into two regimes: the charge-transfer regime ($\Delta<U$) \cite{bocquet1992electronic} and the Mott-Hubbard regime ($\Delta>U$) \cite{zaanen1985band}. However, vdW magnets host small or even negative charge-transfer energy due to the relatively closer positioning of ligand $p$ bands to transition-metal $d$ bands compared to those in oxides. This change in $\Delta$ requires a more comprehensive picture, namely, the configuration interaction theory developed in the 1980s \cite{de1984exciton}. This small charge-transfer effect is crucial for a proper understanding of magnetic exciton physics in NiPS$_3$ \cite{kang2020coherent}. Moreover, charge transfer can also play a crucial role at the interface of a heterostructure, as observed for Fe$_3$GeTe$_2$/Cr$_2$Ge$_2$Te$_6$/Fe$_3$GeTe$_2$, where coercive fields of the ferromagnetic layers could be selectively controlled by the electric-field-assisted directional charge transfer effect \cite{wang2023selectively}. Although this small or negative charge transfer physics has not been thoroughly examined for other vdW magnets and heterostructures, one should not be surprised to see more fundamental changes to our view of vdW magnets in the future that arise from this.

\section{Conclusions}

A simple question about the possibility of "magnetic graphene'', which was asked in the early 2010s, is the primary driver of the vdW magnets. Since then, they have been extensively studied and constitute an important part of contemporary condensed matter physics research. The major advances enabled by vdW magnets include, but are not limited to, the experimental realization of all three classical spin models (Ising, XY, and Heisenberg) in 2D, gate- and twist-tunable magnetism, the creation and control of non-equilibrium magnetic states using light-matter interactions, and finally, the possibility of their use in practical spintronics applications. We believe that future studies of vdW antiferromagnets will lead to a new phase of the spintronic revolution, given their ultrafast spin dynamics and near-zero stray magnetic fields. However, unlocking the full potential of these magnets requires key technological advances, especially in characterization techniques.

\section*{Acknowledgements}
We acknowledge Beom Hyun Kim, Kaixuan Zhang, and Giung Park for their helpful discussions and critical comments. We also thank past and present members of the groups for their work, which provides the basis of this review. Last but not least, we are grateful to our collaborators, who, at various stages of our work, have helped shape much of this review. The work was supported by the Leading Researcher Program of the National Research Foundation of Korea (Grant No. RS-2020-NR049405).

\bibliographystyle{elsarticle-num} 
\bibliography{ref}

\end{document}